# Matrix genetics, part 5: genetic projection operators and direct sums

Sergey V. Petoukhov

Department of Biomechanics, Mechanical Engineering Research Institute of the Russian Academy of Sciences

petoukhov@hotmail.com, petoukhov@imash.ru, http://symmetry.hu/isabm/petoukhov.html

**Corresponding address**: Kutuzovskiy prospect, d. 1/7, kv.58, Moscow, 121248, Russia.

**Abstract**. The article is devoted to phenomena of symmetries and algebras in matrix presentations of the genetic code. The Kronecker family of the genetic matrices is investigated, which is based on the alphabetical matrix [C A; U G], where C, A, U, G are the letters of the genetic alphabet. The matrix P=[C A; U G] in the third Kronecker power is the (8*8)-matrix, which contains 64 triplets. Peculiarities of the degeneracy of the genetic code are reflected in the symmetrical black-and-white mosaic of this genetic (8*8)-matrix of 64 triplets. Phenomena of connections of this mosaic matrix (and many other genetic matrices) with projection operators are revealed. Taking into account an important role of projection operators in quantum mechanics, theory of digital codes, computer science, logic and in many other fields of applied mathematics, we study algebraic properties and biological meanings of these phenomena. Using of notions and formalisms of theory of finite-dimensional vector spaces in bioinformatics and theoretical biology is proposed on the bases of the described results.

## 1 Introduction

Atomic and molecular levels of matter obey the laws of quantum mechanics. Molecular systems of the genetic code are also managed by these laws. It is important to search a connection of formalisms of quantum mechanics with ensembles of molecular structures of the genetic code. It is known that special kinds of matrix operators play important role in quantum mechanics. Taking this into account, one of the main problems of matrix genetics is a detecting of a connection between genetic code structures and matrix formalisms of quantum mechanics.

This article is devoted to a description of connections of molecular-genetic systems with a famous notion of matrix projection operators. These connections were revealed by the author in a course of researches in the field of matrix genetics which studies matrix presentations of the genetic code, first of all, Kronecker families of genetic multiplets [Petoukhov, Petoukhov, He, 2009]. This article continues a description of genetic projection operators which was begun in the article (Petoukhov, arXiv:0803.0888, version 3).

Projection operators are well-known in quantum mechanics, theory of digital codes, computer science, logic and in many other fields of mathematics, mathematical natural sciences and technologies. They are effective tool of mathematical investigations in theory of direct sums of vector spaces, modules, etc. The discovery of their connections with genetic matrices allows transferring relevant ideas and methods of the mathematical sciences into the field of bioinformatics and theoretical biology.

## 2 Some mathematical properties of projection operators

Many properties of projection operators (or projectors, or projections) in linear algebra and functional analysis are described in (Dunford, Schwartz, 1958; Halmos, 1974; Meyer, 2000; http://en.wikipedia.org/wiki/Projection_%28linear_algebra%29 ).

In these mathematical fields a projection operator is a linear transformation $P$ from a vector space to itself such that $P^2 = P$. It is idempotent operator and it leaves its image unchanged. This definition of "projection" formalizes and generalizes the idea of graphical projection. A construction of a plane on the base of two coordinate axes is an analogue of a formation of a direct sum. In this simplest case two types of one-dimensional vectors (x) and (y) of coordinate axes exist. Their combination into two-dimensional vector (x,y) with a certain order of components (x) and (y) leads to a construction of a two-dimensional plane as a direct sum of these one-dimensional spaces.

Let $W$ be a vector space. Suppose the subspaces $U$ and $V$ are the range and null space of $P$ respectively. Then we have these basic properties:

1. $P$ is the identity operator $I$ on $U$: $\forall x \in U : Px = x$.
2. We have a direct sum $W = U \oplus V$. This means that every vector $x$ may be decomposed uniquely in the manner $x = u + v$, where $u$ is in $U$ and $v$ is in $V$. The decomposition is given by $u = Px, \; v = x - Px$.

The range and kernel of a projection are complementary: $P$ and $Q = I - P$. The operator $Q$ is also a projection and the range and kernel of P become the kernel and range of Q and vice-versa. We say $P$ is a projection along $V$ onto $U$ (kernel/range) and $Q$ is a projection along $U$ onto $V$. Decomposition of a vector space into direct sums is not unique in general. Therefore, given a subspace $V$, in general there are many projections whose range (or kernel) is $V$. Only 1 and 0 can be an eigenvalues of a projection. The corresponding eigenspaces are the range and kernel of the projection.

Projection operators are a powerful algebraic tool of studying geometric notion of a direct sum (of vector spaces, of modules, etc.). Study of projections is equivalent to the study of expansions in the direct sums. Studies of algebraic concepts of invariants and reducibility can be conducted by means of projection operators (Halmos, 1974, paragraphs 18, 41 and 43).

Some examples of projections in forms of (n*n)-matrices in multi-dimensional vector spaces are the following:

$$2^{-1} * \begin{vmatrix} +1 & -1 \\ -1 & +1 \end{vmatrix}; \quad (a+b)^{-1} * \begin{vmatrix} a & b \\ a & b \end{vmatrix}; \quad (a+b)^{-1} * \begin{vmatrix} a & a \\ b & b \end{vmatrix}$$

$$(x_0+x_1+\ldots+x_n)^{-1} * \begin{vmatrix} x_0 & x_1 & \ldots & x_n \\ x_0 & x_1 & \ldots & x_n \\ \ldots & \ldots & \ldots & \ldots \\ x_0 & x_1 & \ldots & x_n \end{vmatrix}; \quad (x_0+x_1+\ldots+x_n)^{-1} * \begin{vmatrix} x_0 & x_0 & \ldots & x_0 \\ x_1 & x_1 & \ldots & x_1 \\ \ldots & \ldots & \ldots & \ldots \\ x_n & x_n & \ldots & x_n \end{vmatrix}$$

## 3 A representation of the degeneracy of the genetic code in genetic matrices

Modern science knows many variants (or dialects) of the genetic code, data about which are shown on the NCBI's website http://www.ncbi.nlm.nih.gov/Taxonomy/Utils/wprintgc.cgi. 17 variants (or dialects) of the genetic code exist which differ one from another by some details of correspondences between triplets and objects encoded by them. Most of these dialects (including the so called Standard Code and the Vertebrate Mitochondrial Code) have the following general scheme of their degeneracy where 32 "black" triplets with "strong roots" and 32 "white" triplets with "weak roots" exist.

In this general or basic scheme, the set of 64 triplets contains 16 subfamilies of triplets, every one of which contains 4 triplets with the same two letters on the first positions of each triplet (an example of such subsets is the case of the four triplets CAC, CAA, CAU, CAG with the same two letters CA on their first positions). We shall name such subfamilies as the subfamilies of *NN*-triplets. In the described basic scheme of degeneracy, the set of these 16 subfamilies of *NN*-triplets is divided into two equal subsets from the viewpoint of properties of the degeneracy of the code (Figure 1). The first subset contains 8 subfamilies of so called "two-position" *NN*-triplets, a coding value of which is independent of a letter on their third position. An example of such subfamilies is the four triplets CGC, CGA, CGU, CGC (Figure 1), all of which encode the same amino acid Arg, though they have different letters on their third position. All members of such subfamilies of *NN*-triplets are marked by black color in Figures 1 and 2.

The second subset contains 8 subfamilies of "three-position" NN-triplets, the coding value of which depends on a letter on their third position. An example of such subfamilies in Figure 1 is the four triplets CAC, CAA, CAU, CAC, two of which (CAC, CAU) encode the amino acid His and the other two (CAA, CAG) encode another amino acid Gln. All members of such subfamilies of NN-triplets are marked by white color in Figures 1 and 2. So the whole set of 64 triplets contains two subsets of 32 black triplets and 32 white triplets.

Here one should recall the work (Rumer, 1968) where a combination of letters on the two first positions of each triplet was named as a "root" of this triplet. A set of 64 triplets contains 16 possible variants of such roots. Taking into account properties of triplets, Rumer has divided the set of 16 possible roots into two subsets with eight roots in each. Roots CC, CU, CG, AC, UC, GC, GU, GG form the first of such octets. They were named by Rumer "strong roots". The other eight roots CA, AA, AU, AG, UA, UU, UG, GA form the second octet and they were named weak roots. When Rumer published his works, the Vertebrate Mitochondrial code and some of the other code dialects were unknown. But one can check easily that the set of 32 black (white) triplets, which we show on Figure 3 for cases of the Standard code and the Vertebrate Mitochondrial Code, is identical to the set of 32 triplets with strong (weak) roots described by Rumer. So, using notions proposed by Rumer, the black triplets can be named as triplets with the strong roots and the white triplets can be named as triplets with the weak roots. Rumer believed that this symmetrical division into two binary-oppositional categories of roots is very important for understanding the nature of the genetic code systems.

One can check easily on the basis of data from the mentioned NCBI's website that the following 11 dialects of the genetic code have the same basic scheme of degeneracy with 32 black triplets and with 32 white triplets: 1) the Standard Code; 2) the Vertebrate Mitochondrial Code; 3) the Yeast Mitochondrial Code; 4) the Mold, Protozoan, and Coelenterate Mitochondrial Code and the Mycoplasma/Spiroplasma Code; 5) the Ciliate, Dasycladacean and Hexamita Nuclear Code; 6) the Euplotid Nuclear Code; 7) the Bacterial and Plant Plastid Code; 8) the Ascidian Mitochondrial Code; 9) the Blepharisma Nuclear Code; 10) the Thraustochytrium Mitochondrial Code; 11) the Chlorophycean Mitochondrial Code. In this article we will consider this basic scheme of the degeneracy which is presented by means of a black-and-white mosaics in a family of genetic matrices ([C A; U G]$^{(3)}$, etc.) on Figure 2.

One can mentioned that the other 6 dialects of the genetic code have only small differences from the described basic scheme of the degeneracy: the Invertebrate Mitochondrial Code; the Echinoderm and Flatworm Mitochondrial Code; the Alternative Yeast Nuclear Code; the Alternative Flatworm Mitochondrial Code; the Trematode Mitochondrial Code; the Scenedesmus obliquus mitochondrial Code.

According to general traditions, the theory of symmetry studies initially those natural objects which possess the most symmetrical character, and then it constructs a theory for cases of violations of this symmetry in other kindred objects. For this reason one should pay special attention to the Vertebrate Mitochondrial code which is the most symmetrical code among dialects of the genetic code and which corresponds to the basic scheme of the degeneracy. One can mention additionally that some authors consider this dialect not only as the most "perfect" but also as the most ancient dialect (Frank-Kamenetskiy, 1988), but this last aspect is a debatable one. Figure 1 shows the correspondence between the set of 64 triplets and the set of 20 amino acids with stop-signals (Stop) of protein synthesis in the Standard Code and in the Vertebrate Mitochondrial Code.

| THE STANDARD CODE ||
|---|---|
| 8 subfamilies of the "two-position NN-triplets" ("black triplets") and the amino acids, which are encoded by them | 8 subfamilies of the "three-position NN-triplets" ("white triplets") and the amino acids, which are encoded by them |
| CCC, CCU, CCA, CCG ➔ Pro | CAC, CAU, CAA, CAG ➔ His, His, Gln, Gln |
| CUC, CUU, CUA, CUG ➔ Leu | AAC, AAU, AAA, AAG ➔ Asn, Asn, Lys, Lys |
| CGC, CGU, CGA, CGG ➔ Arg | AUC, AUU, AUA, AUG ➔ Ile, Ile, Ile, Met |
| ACC, ACU, ACA, ACG ➔ Thr | AGC, AGU, AGA, AGG ➔ Ser, Ser, Arg, Arg |
| UCC, UCU, UCA, UCG ➔ Ser | UAC, UAU, UAA, UAG ➔ Tyr, Tyr, Stop, Stop |
| GCC, GCU, GCA, GCG ➔ Ala | UUC, UUU, UUA, UUG ➔ Phe, Phe, Leu, Leu |
| GUC, GUU, GUA, GUG ➔ Val | UGC, UGU, UGA, UGG ➔ Cys, Cys, Stop, Trp |
| GGC, GGU, GGA, GGG ➔ Gly | GAC, GAU, GAA, GAG ➔ Asp, Asp, Glu, Glu |

| THE VERTEBRATE MITOCHONDRIAL CODE ||
|---|---|
| 8 subfamilies of the "two-position NN-triplets" ("black triplets") and the amino acids, which are encoded by them | 8 subfamilies of the "three-position NN-triplets" ("white triplets") and the amino acids, which are encoded by them |
| CCC, CCU, CCA, CCG ➔ Pro | CAC, CAU, CAA, CAG ➔ His, His, Gln, Gln |
| CUC, CUU, CUA, CUG ➔ Leu | AAC, AAU, AAA, AAG ➔ Asn, Asn, Lys, Lys |
| CGC, CGU, CGA, CGG ➔ Arg | AUC, AUU, AUA, AUG ➔ Ile, Ile, Met, Met |
| ACC, ACU, ACA, ACG ➔ Thr | AGC, AGU, AGA, AGG ➔ Ser, Ser, Stop, Stop |
| UCC, UCU, UCA, UCG ➔ Ser | UAC, UAU, UAA, UAG ➔ Tyr, Tyr, Stop, Stop |
| GCC, GCU, GCA, GCG ➔ Ala | UUC, UUU, UUA, UUG ➔ Phe, Phe, Leu, Leu |
| GUC, GUU, GUA, GUG ➔ Val | UGC, UGU, UGA, UGG ➔ Cys, Cys, Trp, Trp |
| GGC, GGU, GGA, GGG ➔ Gly | GAC, GAU, GAA, GAG ➔ Asp, Asp, Glu, Glu |

*Figure 1. Two examples of the basic scheme of the genetic code degeneracy with 32 "black" triplets and 32 "white" triplets. Top: the case of the Standard Code. Bottom: the case of the Vertebrate Mitochondrial Code. All initial data are taken from the NCBI's web-site http://www.ncbi.nlm.nih.gov/Taxonomy/Utils/wprintgc.cgi.*

Let us return to studying the genetic matrix $[C\ A;\ U\ G]^{(3)}$ where C, A, G, U (cytosine, adenine, uracil, guanine) are the letters of the genetic alphabet, (3) means the third Kronecker exponentiation (initial researches of such genetic matrices were described in (Petoukhov, 2001a, 2006, 2008a; Petoukhov, He, 2009). This formally constructed matrix contains all 64 triplets in a strong order (Figure 2, the upper matrix).

The author has revealed an unexpected phenomenological fact of a very symmetrical disposition of black triplets and white triplets in the genomatrix $[C\ A;\ U\ G]^{(3)}$, which was constructed formally without any mention about amino acids and the degeneracy of the genetic code (Figure 2). In the result this genomatrix has a certain black-and-white mosaic which reflects features of the basic scheme of the degeneracy of the genetic code. One should emphasize that the most of possible variants of dispositions of black triplets and white triplets give quite asymmetric mosaics of halves and quadrants of 8*8-matrices (the total quantity of these variants for 8*8-matrix is equal to huge number $64!=10^{89}$).

It is unexpected also that any kind of permutation of positions in triplets, which is accomplished in all 64 triplets simultaneously, leads to the transformed genomatrix, which possesses a symmetrical black-and-white mosaic of degeneracy also. The six kinds of sequences of positions in triplets exist: 1-2-3, 2-3-1, 3-1-2, 1-3-2, 2-1-3, 3-2-1. It is obvious that if the positional sequence 1-2-3 in triplets is replaced, for example, by the sequence 2-3-1, the most triplets change their disposition in the genomatrix. In this case the initial genomatrix is reconstructed cardinally into the new mosaic matrix. For instance, in the result of such permutation the black triplet CGA is replaced in its matrix cell by the white triplet GAC, etc. Let us denote the six genomatrices, which correspond to the mentioned kinds of positional sequences in triplets, by the symbols $P^{CAUG}_{123}(=P^{(3)})$, $P^{CAUG}_{231}$, $P^{CAUG}_{312}$, $P^{CAUG}_{132}$, $P^{CAUG}_{213}$, $P^{CAUG}_{321}$. Here the bottom indexes show the appropriate positional sequences in triplets; the upper index shows the kind of basic alphabetical matrix $[C\ A;\ U\ G]$ of the Kronecker family (later we shall consider other cases of such basic genomatrices). Figure 2 demonstrates these six genomatrices.

Each of these six genomatrices has a symmetric character unexpectedly (Figure 2). For example, the first genomatrix $P^{CAUG}_{123} = [C\ A;\ U\ G]^{(3)}$ has the following symmetric features:

1. The left and right halves of this matrix are mirror-antisymmetric to each other by its colors: any pair of cells, disposed by the mirror-symmetric manner in these halves, has opposite colors.
2. Diagonal quadrants of the matrix are identical to each other from the viewpoint of their mosaics.
3. The adjacent rows 0-1, 2-3, 4-5, 6-7 are identical to each other from the viewpoint of the mosaic and of the disposition of the same amino acids in their proper cells.
4. Mosaics of all rows of the (8x8)-genomatrix and of its (4x4)-quadrants have a meander-line character, which is connected with Rademacher functions from theory of digital signal processing.

Each of the other 5 genomatrices on Figure 4 has symmetrical properties also, first of all, the following two properties:

1. The left and right halves of each matrix are mirror-antisymmetric to each other by its colors: any pair of cells, disposed by the mirror-symmetric manner in these halves, has opposite colors.
2. Mosaics of all rows of each (8x8)-genomatrix and of its (4x4)-quadrants have a meander-line character, which is connected with Rademacher functions from theory of digital signal processing.

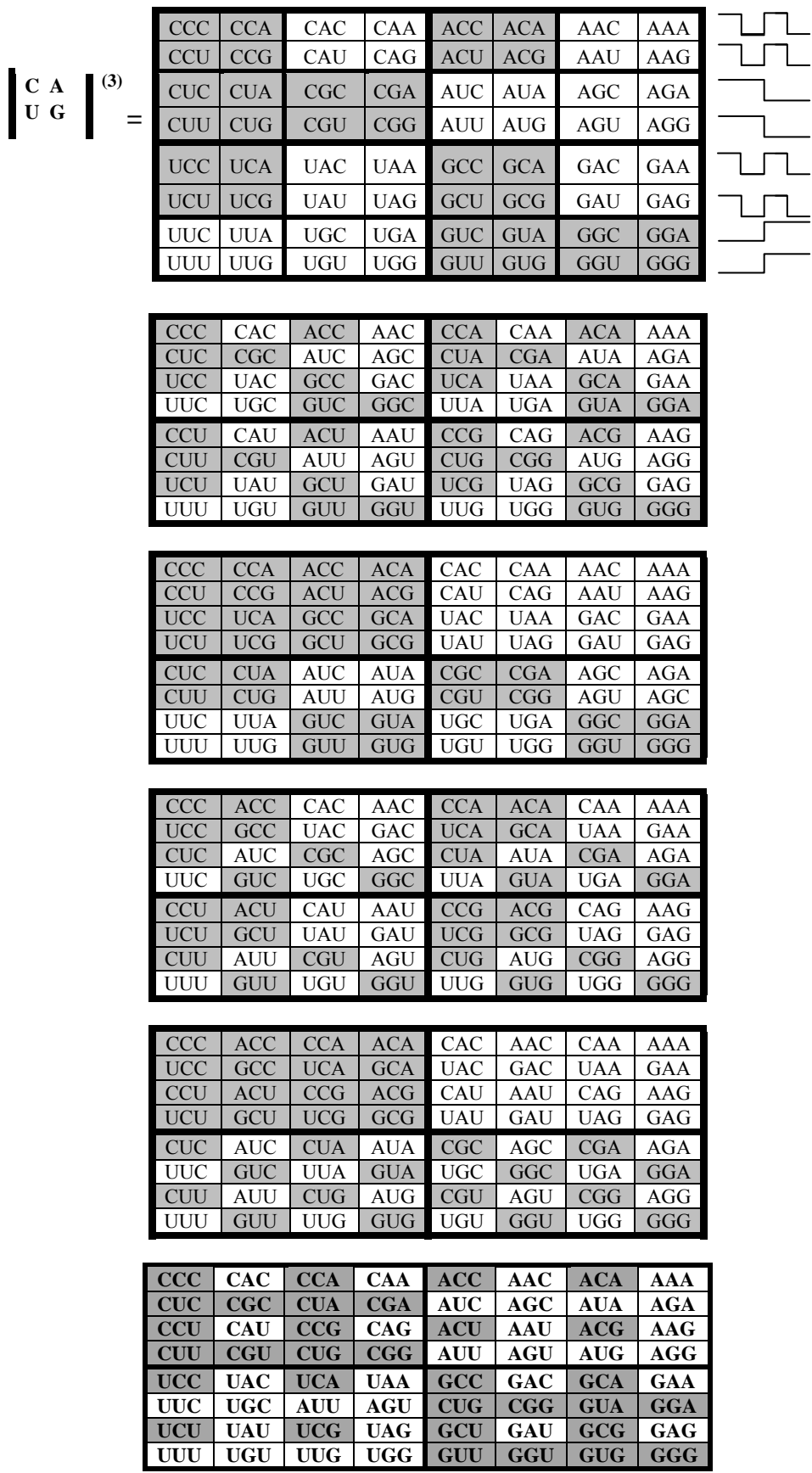

Figure 2. The genomatrices $P^{CAUG}_{123}= [C\ A;\ U\ G]^{(3)}$, $P^{CAUG}_{231}$, $P^{CAUG}_{213}$, $P^{CAUG}_{321}$, $P^{CAUG}_{312}$, $P^{CAUG}_{132}$. Each matrix cell has a triplet and an amino acid (or stop-signal) coded by this triplet. The black-and-white mosaic presents a specificity of the basic scheme of genetic code degeneracy. Rademacher functions are shown here only for every row of the upper matrix $[C\ A;\ U\ G]^{(3)}$.

Rademacher functions are described by the expression $r_n(t) = \text{sign}(\sin 2^n \pi t)$, $n = 1, 2, 3, ...$ (here sign is a function of sign of the argument). Sometimes such functions are called "square waves" (http://mathworld.wolfram.com/RademacherFunction.html). Rademacher functions take only two values "+1" and "-1" (Figure 3). These functions form an incomplete orthogonal system of functions, each of which is an odd function (that defines the mirror antisymmetry between the left and right halves of genomatrices). Rademacher functions are connected to a complete orthogonal system of Walsh functions and Hadamard matrices and to Gray code also.

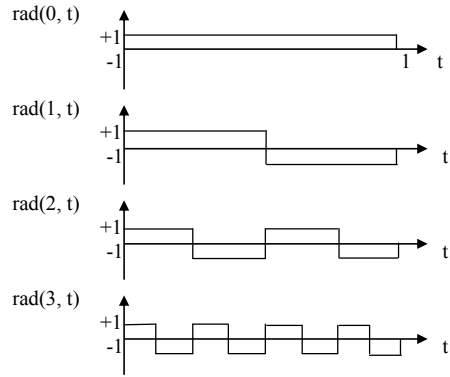

Figure 3. Examples of Rademacher functions.

It seems essential that this connection between the matrix rows (or columns in some cases) and Rademacher functions is a conserved stability in a great number of new genomatrices which are produced by means of positional and alphabetic permutations of genetic elements inside triplets in the initial genomatrix $P^{CAUG}_{123} = [C\ A;\ U\ G]^{(3)}$ (Petoukhov, 2006, 2008a-c; Petoukhov, He, 2009).

## 3  Projection operators and genetic matrices

Taking this fact of the stable connection of the genetic matrices with Rademacher functions into account, let us represent the black-and-white mosaic of each from the mentioned six genomatrices as a binary mosaic of numbers "+1" and "-1" by means of replacing black (white) color of each matrix cell by an element "+1" ("-1"). In such "Rademacher presentation", the genomatrices $P^{CAUG}_{123}$, $P^{CAUG}_{231}$, $P^{CAUG}_{213}$, $P^{CAUG}_{321}$, $P^{CAUG}_{312}$, $P^{CAUG}_{132}$ are reformed into the genomatrices $B_{123}$, $B_{231}$, $B_{312}$, $B_{132}$, $B_{213}$, $B_{321}$ (Figure 4).

An unexpected general property of these six binary genomatrices on Figure 4 is their connection with projection operators. Really let us consider matrices $Y_{123} = 4^{-1} * B_{123}$, $Y_{231} = 4^{-1} * B_{231}$, $Y_{312} = 4^{-1} * B_{312}$, $Y_{321} = 4^{-1} * B_{321}$, $Y_{213} = 4^{-1} * B_{213}$, $Y_{132} = 4^{-1} * B_{132}$. One can check that each of this matrices $Y_k$ is a projection operator which satisfies the defining condition $P^2 = P$:

$$(Y_{123})^2 = Y_{123}; \quad (Y_{231})^2 = Y_{231}; \quad (Y_{312})^2 = Y_{312}$$

$$(Y_{132})^2 = Y_{132}; \quad (Y_{213})^2 = Y_{213}; \quad (Y_{321})^2 = Y_{321} \qquad (1)$$

We will name projector operators, which are based on different kinds of genetic matrices, as "genoprojector operators" (or "genoprojectors"). Concerning to the 8*8-matrices $B_{123}$, $B_{231}$, $B_{312}$, $B_{132}$, $B_{213}$, $B_{321}$ (Figure 4), one can add that their 4*4-quadrants and 2*2-subquadrants are connected in many cases (not in all cases) with projection operators also but with different scaling coefficients.

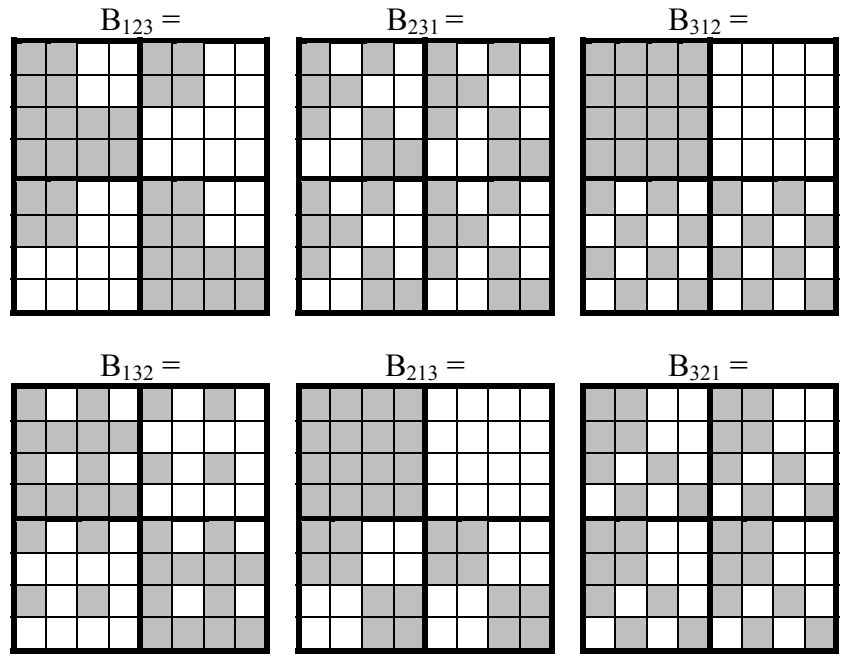

Figure 4. Numeric genomatrices $B_{123}$, $B_{231}$, $B_{312}$, $B_{132}$, $B_{213}$, $B_{321}$, in which each black cell means the element "+1" and each white cell means the element "-1".

Some of these genoprojectors are commutative, other are non-commutative. Figure 5 shows examples of pairs of commutative operators whose indices differ in reverse order reading. Of course, each of new matrices $Y_{123}*Y_{321}$, $Y_{231}*Y_{132}$, $Y_{312}*Y_{213}$ is a projector also.

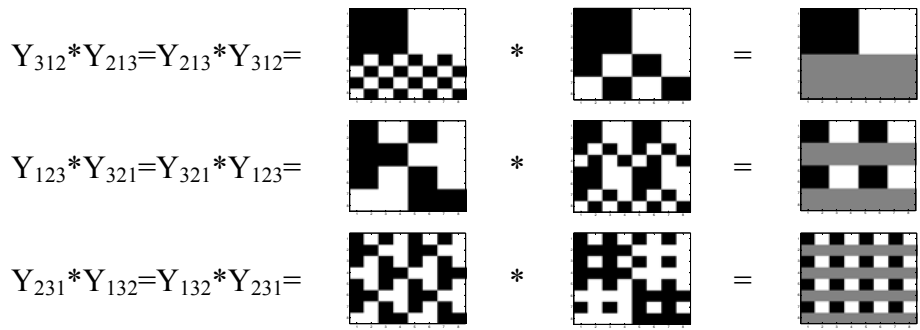

Figure 5. Pairs of (8*8)-genomatrices of commutative projectors in their visual presentations. Here each black cell corresponds to "+1/4", each white cell – "-1/4", each grey cell – "0".

By definition, two projectors $P_1$ and $P_2$ are named orthogonal if $P_1*P_2=0$. One can check that the three genoprojectors $Y_{312}*Y_{213}$, $Y_{123}*Y_{321}$, $Y_{231}*Y_{132}$ (Figure 8) are orthogonal each to other. It is known that in a case of two orthogonal projectors $P_1$ and $P_2$, which satisfy a condition $P_1*P_2 = P_2*P_1 = 0$, their sum $P_1+P_2$ is a projector also (Halmos, 1974, paragraph 42).

One should mention additionally that in the case of the Vertebrate Mitochondrial Code (Figure 1), which is the most symmetrical among all variants of the genetic codes, all amino acids and stop-codons are disposed phenomenologically in the set of 2*2-subquadrants of the 8*8-genomatrix $[C\ A; U\ G]^{(3)}$ in such way that each of the subquadrants corresponds to the scheme of a projector $(a+b)^{-1}*[a\ b;\ a\ b]$ (see Figure 6). For example, if one takes quantities of protons in molecules of amino acids His and Gln, which are disposed in one subquadrant, then this subquadrant corresponds to the projector $160^{-1}*[82\ 78;\ 82\ 78]$, where 82 and 78 are the quantities of protons in His and Gln correspondingly. From such viewpoint, this 8*8-genomatrix is a system of 2*2-matrices of projection operators.

| CCC<br>Pro | CCA<br>Pro | CAC<br>His | CAA<br>Gln | ACC<br>Thr | ACA<br>Thr | AAC<br>Asn | AAA<br>Lys |
|---|---|---|---|---|---|---|---|
| CCU<br>Pro | CCG<br>Pro | CAU<br>His | CAG<br>Gln | ACU<br>Thr | ACG<br>Thr | AAU<br>Asn | AAG<br>Lys |
| CUC<br>Leu | CUA<br>Leu | CGC<br>Arg | CGA<br>Arg | AUC<br>Ile | AUA<br>Met | AGC<br>Ser | AGA<br>Stop |
| CUU<br>Leu | CUG<br>Leu | CGU<br>Arg | CGG<br>Arg | AUU<br>Ile | AUG<br>Met | AGU<br>Ser | AGG<br>Stop |
| UCC<br>Ser | UCA<br>Ser | UAC<br>Tyr | UAA<br>Stop | GCC<br>Ala | GCA<br>Ala | GAC<br>Asp | GAA<br>Glu |
| UCU<br>Ser | UCG<br>Ser | UAU<br>Tyr | UAG<br>Stop | GCU<br>Ala | GCG<br>Ala | GAU<br>Asp | GAG<br>Glu |
| UUC<br>Phe | UUA<br>Leu | UGC<br>Cys | UGA<br>Trp | GUC<br>Val | GUA<br>Val | GGC<br>Gly | GGA<br>Gly |
| UUU<br>Phe | UUG<br>Leu | UGU<br>Cys | UGG<br>Trp | GUU<br>Val | GUG<br>Val | GGU<br>Gly | GGG<br>Gly |

Figure 6. The disposition of amino acids and stop-codons inside the genomatrix [C A; U G]$^{(3)}$ in the case of the Vertebrate Mitochondrial Code.

One can mention else about a possible genetic meaning of the scaling coefficient 4 which connects the matrices $B_{jkl}$ and $Y_{jkl}$ from the expressions (1). The matrices $B_{jkl}$ satisfy to the following condition: $(B_{jkl})^2 = 4*B_{jkl}$. This mathematical property generates heuristic associations with a famous genetic phenomenon of a tetra-reproduction of gametal cells in a course of meiosis which are carriers of genetic information. One can think that such "tetra-scaling projectors" $B_{jkl}$ can be used in a mathematical simulation of meiosis, etc.

The desribed (8*8)-matrices of genoprojectors $Y_{123}$, $Y_{231}$, $Y_{312}$, $Y_{321}$, $Y_{213}$, $Y_{132}$ are connected with genetic 8-dimensional Yin-Yang algebras (or bipolar algebras) which were presented in our works (Petoukhov, arXiv:0803.3330 and arXiv:0805.4692; Petoukhov, He, 2009). For example the genoprojector $Y_{123}$ from the expression (1) can be presented in a form: $Y_{123} = 4^{-1}*(f_0 + m_1 + f_2 + m_3 + f_4 + m_5 + f_6 + m_7)$ where $f_0$, $m_1$, $f_2$, $m_3$, $f_4$, $m_5$, $f_6$, $m_7$ are the basic matrices of the corresponding 8-dimensional Yin-Yang algebra (see Figure 4 in the article (Petoukhov, arXiv:0803.3330)). In other words the genoprojector $Y_{123}$ is decomposed into sum of these basic matrices of Yin-type ($f_0$, $f_2$, $f_4$, $f_6$) and of Yang-type ($m_1$, $m_3$, $m_5$, $m_7$). The matrices $f_0$ and $m_1$ are genoprojectors each of which are sum of four orthogonal projectors, etc. Such analysis of genetic Yin-Yang algebras from the viewpoint of projection operators will be published later together with our data about a connection of genetic projectors with biological evolution of dialects of the genetic code.

We calculated eigenvectors and eigenvalues of the genoprojectors $Y_{123}$, $Y_{231}$, $Y_{312}$, $Y_{321}$, $Y_{213}$, $Y_{132}$ from the expressions (1) by means of the computer program MatLab. The result is that in each of these genoprojectors, two eigenvalues are equal to "1" and other six eigenvalues are equal to "0". Figure 7 shows eigenvectors of these genoprojectors. Those eigenvectors which correspond to eigenvalues „1" are highlighted by yellow.

|  | Eigenvectors | | | | | | | |
|---|---|---|---|---|---|---|---|---|
| **Y₁₂₃** | -0.7500 | **8⁻⁰·⁵** | 0.1581 | **-0.2784** | 0.7735 | -0.2438 | -0.3756 | -0.0000 |
|  | 0.2500 | **8⁻⁰·⁵** | 0.5248 | **-0.2784** | -0.0073 | -0.4462 | -0.2495 | 0.0000 |
|  | 0.2500 | **8⁻⁰·⁵** | -0.3415 | **0.4153** | -0.2291 | 0.3825 | 0.3126 | 0.0000 |
|  | 0.2500 | **8⁻⁰·⁵** | -0.3415 | **0.4153** | -0.0830 | 0.3258 | 0.3126 | 0.0000 |
|  | 0.2500 | **8⁻⁰·⁵** | -0.3415 | **-0.2784** | -0.1560 | 0.3542 | -0.0119 | 0.0000 |
|  | 0.2500 | **8⁻⁰·⁵** | -0.3415 | **-0.2784** | -0.1560 | 0.3542 | 0.6370 | 0.0000 |
|  | -0.2500 | **-8⁻⁰·⁵** | -0.3415 | **-0.4153** | 0.3831 | -0.3450 | -0.3126 | -0.7071 |
|  | -0.2500 | **-8⁻⁰·⁵** | 0.3415 | **-0.4153** | 0.3831 | -0.3450 | -0.3126 | 0.7071 |
| **Y₂₃₁** | -0.7500 | **8⁻⁰·⁵** | -0.7500 | **0.3296** | -0.1360 | -0.3274 | -0.0829 | 0.3061 |
|  | 0.2500 | **8⁻⁰·⁵** | 0.2500 | **-0.3760** | 0.6476 | 0.4671 | -0.6601 | -0.7004 |
|  | 0.2500 | **8⁻⁰·⁵** | 0.2500 | **0.3296** | 0.4854 | 0.1887 | -0.6837 | -0.5281 |
|  | -0.2500 | **-8⁻⁰·⁵** | -0.2500 | **0.3760** | -0.4494 | -0.6676 | 0.1125 | 0.2723 |
|  | 0.2500 | **8⁻⁰·⁵** | 0.2500 | **0.3296** | -0.0756 | -0.0309 | 0.0574 | -0.0087 |
|  | 0.2500 | **8⁻⁰·⁵** | 0.2500 | **-0.3760** | -0.2378 | -0.3093 | 0.1534 | 0.2396 |
|  | 0.2500 | **8⁻⁰·⁵** | 0.2500 | **0.3296** | -0.0756 | -0.0309 | 0.1770 | 0.0673 |
|  | -0.2500 | **-8⁻⁰·⁵** | -0.2500 | **0.3760** | 0.2378 | 0.3093 | -0.1380 | 0.0251 |
| **Y₃₁₂** | -0.7500 | **8⁻⁰·⁵** | 0.1992 | **0.3995** | -0.1924 | -0.5201 | 0.2507 | -0.2688 |
|  | 0.2500 | **8⁻⁰·⁵** | -0.7582 | **0.3995** | 0.2591 | -0.0167 | -0.1733 | 0.1117 |
|  | 0.2500 | **8⁻⁰·⁵** | 0.2795 | **0.3995** | 0.5772 | -0.3186 | 0.3919 | -0.2105 |
|  | 0.2500 | **8⁻⁰·⁵** | 0.2795 | **0.3995** | -0.6439 | 0.1677 | 0.0352 | 0.0290 |
|  | 0.2500 | **8⁻⁰·⁵** | -0.2394 | **-0.3007** | -0.1924 | -0.2684 | -0.5735 | 0.4090 |
|  | -0.2500 | **-8⁻⁰·⁵** | 0.2394 | **0.3007** | 0.1924 | -0.4194 | 0.4383 | -0.7476 |
|  | 0.2500 | **8⁻⁰·⁵** | -0.2394 | **-0.3007** | -0.1924 | 0.4194 | 0.4353 | -0.2683 |
|  | -0.2500 | **-8⁻⁰·⁵** | 0.2394 | **0.3007** | 0.1924 | -0.4194 | 0.2042 | 0.2683 |
| **Y₃₂₁** | -0.7500 | **8⁻⁰·⁵** | **-0.3713** | 0.5335 | 0.5335 | 0.0492 | 0.2908 | -0.1715 |
|  | 0.2500 | **8⁻⁰·⁵** | **-0.3713** | 0.0608+0.1184i | 0.0608-0.1184i | 0.4480 | -0.1911 | -0.3452 |
|  | 0.2500 | **8⁻⁰·⁵** | **0.3349** | 0.2612-0.0339i | 0.2612+0.0339i | 0.0293 | 0.2723 | -0.1425 |
|  | -0.2500 | **-8⁻⁰·⁵** | **-0.3349** | -0.1923+0.0224i | -0.1923-0.0224i | 0.5071 | -0.3779 | -0.5987 |
|  | 0.2500 | **8⁻⁰·⁵** | **-0.3713** | -0.2627-0.0650i | -0.2627+0.0650i | -0.0554 | -0.1026 | 0.1995 |
|  | 0.2500 | **8⁻⁰·⁵** | **-0.3713** | -0.2627-0.0650i | -0.2627+0.0650i | 0.0945 | -0.1026 | 0.0021 |
|  | 0.2500 | **8⁻⁰·⁵** | **0.3349** | -0.4631+0.0874i | -0.4631-0.0874i | 0.5133 | -0.5661 | -0.2005 |
|  | -0.2500 | **-8⁻⁰·⁵** | **-0.3349** | 0.4631-0.0874i | 0.4631+0.0874i | -0.5133 | 0.5661 | 0.6267 |
| **Y₂₁₃** | -0.7500 | **8⁻⁰·⁵** | 0.1610 | **0.0293** | 0.2479 | -0.2690 | 0.2479 | 0.1698 |
|  | 0.2500 | **8⁻⁰·⁵** | 0.5226 | **0.0293** | 0.4388 | -0.4207 | 0.4388 | 0.0020 |
|  | 0.2500 | **8⁻⁰·⁵** | -0.3418 | **0.0293** | -0.2040 | 0.2084 | -0.2040 | -0.0859 |
|  | 0.2500 | **8⁻⁰·⁵** | -0.3418 | **0.0293** | -0.4827 | 0.4812 | -0.4827 | -0.0859 |
|  | 0.2500 | **8⁻⁰·⁵** | -0.3418 | **0.4991** | -0.3433 | 0.3448 | -0.3433 | -0.0859 |
|  | 0.2500 | **8⁻⁰·⁵** | -0.3418 | **0.4991** | -0.3433 | 0.3448 | -0.3433 | -0.0859 |
|  | -0.2500 | **-8⁻⁰·⁵** | 0.3418 | **-0.4991** | 0.3433 | -0.3448 | 0.3433 | -0.5948 |
|  | -0.2500 | **-8⁻⁰·⁵** | 0.3418 | **-0.4991** | 0.3433 | -0.3448 | 0.3433 | 0.7667 |
| **Y₁₃₂** | -0.7500 | **8⁻⁰·⁵** | -0.7500 | **-0.3438** | 0.7885 | -0.0933 | -0.3133 | 0.3395 |
|  | 0.2500 | **8⁻⁰·⁵** | 0.2500 | **0.3630** | 0.1770 | -0.2458 | 0.2632 | -0.5213 |
|  | 0.2500 | **8⁻⁰·⁵** | 0.2500 | **-0.3438** | -0.1165 | -0.5769 | 0.1679 | -0.2764 |
|  | 0.2500 | **8⁻⁰·⁵** | 0.2500 | **0.3630** | -0.3136 | -0.4035 | 0.0727 | -0.0316 |
|  | 0.2500 | **8⁻⁰·⁵** | 0.2500 | **-0.3438** | -0.0683 | -0.3246 | 0.1320 | 0.1599 |
|  | -0.2500 | **-8⁻⁰·⁵** | -0.2500 | **-0.3630** | 0.3360 | -0.3351 | -0.6766 | 0.0316 |
|  | 0.2500 | **8⁻⁰·⁵** | 0.2500 | **-0.3438** | -0.0683 | -0.3246 | 0.2039 | -0.7127 |
|  | -0.2500 | **-8⁻⁰·⁵** | -0.2500 | **-0.3630** | 0.3360 | -0.3351 | 0.5312 | 0.0316 |

Figure 7. Eigenvectors of the genetic projection operators $Y_{123}, Y_{231}, Y_{312}, Y_{321}, Y_{213}, Y_{132}$. Yellow color highlights eigenvectors which correspond to eigenvalues "1".

## 4 Genetic matrices with a diagonal disposition of complementary pairs

Till now we described the case of alphabetical matrix [C A; U G] where the complementary pairs C-G and A-U were disposed along matrix diagonals (Figure 2). But a set of alphabetical matrices exists where these complementary pairs are disposed along matrix diagonals also (Figure 8).

$$\begin{vmatrix} C & A \\ U & G \end{vmatrix}; \begin{vmatrix} C & U \\ A & G \end{vmatrix}; \begin{vmatrix} G & A \\ U & C \end{vmatrix}; \begin{vmatrix} G & U \\ A & C \end{vmatrix}; \begin{vmatrix} A & C \\ G & U \end{vmatrix}; \begin{vmatrix} A & G \\ C & U \end{vmatrix}; \begin{vmatrix} U & C \\ G & A \end{vmatrix}; \begin{vmatrix} U & G \\ C & A \end{vmatrix}$$

Figure 8. Examples of alphabetical genomatrices with a diagonal disposition of complementary pairs C-G and A-U.

The third Kronecker power of each of these alphabetical matrices contains all 64 triplets in an individual order. The phenomenon of the degeneracy of the genetic code generates an individual black-and-white mosaic for every of these 8*8-matrices (in this article we are considering only the basic scheme of the degeneracy of the code with 32 black triplets and 32 white triplets which was described above). Have these new genetic 8*8-matrices a connection with projection operators or not?

The answer is positive: every of these genomatrices [C U; A G]$^{(3)}$, [G A; U C]$^{(3)}$, [G U; A C]$^{(3)}$, etc. has a connection with projection operators which is similar to the connection in the described case of [C A; U G]$^{(3)}$. For example, let us demonstrate this fact for the case of the genomatrix [C U; A G]$^{(3)}$ = $P^{CUAG}_{123}$. Figure 9 shows this genomatrix (the order of positions in its triplets is 1-2-3) and five its modifications $P^{CUAG}_{231}$, $P^{CUAG}_{312}$, $P^{CUAG}_{321}$$^{(3)}$, $P^{CUAG}_{213}$, $P^{CUAG}_{132}$ which are made by means of the permutation of positions inside triplets (2-3-1, 3-1-2, 3-2-1, 2-1-3, 1-3-2). One can see that a mosaic of every column of these matrices corresponds to one of Rademacher functions (in contrast to the previous case in Figure 2 where each row corresponds to one of Rademacher functions).

By analogy with Rademacher presentations of the genomatrices in Figure 4, one can study Rademacher presentations of these genomatrices from Figure 9. In other words, taking into account the fact of the stable connection with Rademacher functions, one can represent the black-and-white mosaic of each from the mentioned six genomatrices (Figure 9) as a mosaic of numbers "+1" and "-1" by means of replacing black (white) color of each matrix cell by an element "+1" ("-1"). In such way the genomatrices $P^{CUAG}_{123}$ = [C U; A G]$^{(3)}$, $P^{CUAG}_{231}$, $P^{CUAG}_{312}$, $P^{CUAG}_{321}$$^{(3)}$, $P^{CUAG}_{213}$, $P^{CUAG}_{132}$ are reformed into the genomatrices $V_{123}$, $V_{231}$, $V_{312}$, $V_{321}$, $V_{213}$, $V_{132}$ (Figure 10).

$$\begin{vmatrix} C & U \\ A & G \end{vmatrix}^{(3)} =$$

| CCC | CCU | CUC | CUU | UCC | UCU | UUC | UUU |
|-----|-----|-----|-----|-----|-----|-----|-----|
| CCA | CCG | CUA | CUG | UCA | UCG | UUA | UUG |
| CAC | CAU | CGC | CGU | UAC | UAU | UGC | UGU |
| CAA | CAG | CGA | CGG | UAA | UAG | UGA | UGG |
| ACC | ACU | AUC | AUU | GCC | GCU | GUC | GUU |
| ACA | ACG | AUA | AUG | GCA | GCG | GUA | GUG |
| AAC | AAU | AGC | AGU | GAC | GAU | GGC | GGU |
| AAA | AAG | AGA | AGG | GAA | GAG | GGA | GGG |

| CCC | CUC | UCC | UUC | CCU | CUU | UCU | UUU |
|-----|-----|-----|-----|-----|-----|-----|-----|
| CAC | CGC | UAC | UGC | CAU | CGU | UAU | UGU |
| ACC | AUC | GCC | GUC | ACU | AUU | GCU | GUU |
| AAC | AGC | GAC | GGC | AAU | AGU | GAU | GGU |
| CCA | CUA | UCA | UUA | CCG | CUG | UCG | UUG |
| CAA | CGA | UAA | UGA | CAG | CGG | UAG | UGG |
| ACA | AUA | GCA | GUA | ACG | AUG | GCG | GUG |
| AAA | AGA | GAA | GGA | AAG | AGG | GAG | GGG |

| CCC | UCC | CCU | UCU | CUC | UUC | CUU | UUU |
|-----|-----|-----|-----|-----|-----|-----|-----|
| ACC | GCC | ACU | GCU | AUC | GUC | AUU | GUU |
| CCA | UCA | CCG | UCG | CUA | UUA | CUG | UUG |
| ACA | GCA | ACG | GCG | AUA | GUA | AUG | GUG |
| CAC | UAC | CAU | UAU | CGC | UGC | CGU | UGU |
| AAC | GAC | AAU | GAU | AGC | GGC | AGU | GGU |
| CAA | UAA | CAG | UAG | CGA | UGA | CGG | UGG |
| AAA | GAA | AAG | GAG | AGA | GGA | AGG | GGG |

| CCC | UCC | CUC | UUC | CCU | UCU | CUU | UUU |
|-----|-----|-----|-----|-----|-----|-----|-----|
| ACC | GCC | AUC | GUC | ACU | GCU | AUU | GUU |
| CAC | UAC | CGC | UGC | CAU | UAU | CGU | UGU |
| AAC | GAC | AGC | GGC | AAU | GAU | AGU | GGU |
| CCA | UCA | CUA | UUA | CCG | UCG | CUG | UUG |
| ACA | GCA | AUA | GUA | ACG | GCG | AUG | GUG |
| CAA | UAA | CGA | UGA | CAG | UAG | CGG | UGG |
| AAA | GAA | AGA | GGA | AAG | GAG | AGG | GGG |

| CCC | CCU | UCC | UCU | CUC | CUU | UUC | UUU |
|-----|-----|-----|-----|-----|-----|-----|-----|
| CCA | CCG | UCA | UCG | CUA | CUG | UUA | UUG |
| ACC | ACU | GCC | GCU | AUC | AUU | GUC | GUU |
| ACA | ACG | GCA | GCG | AUA | AUG | GUA | GUG |
| CAC | CAU | UAC | UAU | CGC | CGU | UGC | UGU |
| CAA | CAG | UAA | UAG | CGA | CGG | UGA | UGG |
| AAC | AAU | GAC | GAU | AGC | AGU | GGC | GGU |
| AAA | AAG | GAA | GAG | AGA | AGG | GGA | GGG |

| CCC | CUC | CCU | CUU | UCC | UUC | UCU | UUU |
|-----|-----|-----|-----|-----|-----|-----|-----|
| CAC | CGC | CAU | CGU | UAC | UGC | UAU | UGU |
| CCA | CUA | CCG | CUG | UCA | UUA | UCG | UUG |
| CAA | CGA | CAG | CGG | UAA | UGA | UAG | UGG |
| ACC | AUC | ACU | AUU | GCC | GUC | GCU | GUU |
| AAC | AGC | AAU | AGU | GAC | GGC | GAU | GGU |
| ACA | AUA | ACG | AUG | GCA | GUA | GCG | GUG |
| AAA | AGA | AAG | AGG | GAA | GGA | GAG | GGG |

Figure 9. The genomatrices $P^{CUAG}_{123}= [C\ U;\ A\ G]^{(3)}$, $P^{CUAG}_{231}$, $P^{CUAG}_{312}$, $P^{CUAG}_{321}{}^{(3)}$, $P^{CUAG}_{213}$, $P^{CUAG}_{132}$. The black-and-white mosaic presents a specificity of the basic scheme of genetic code degeneracy. A mosaic of every column of these matrices corresponds to one of Rademacher functions.

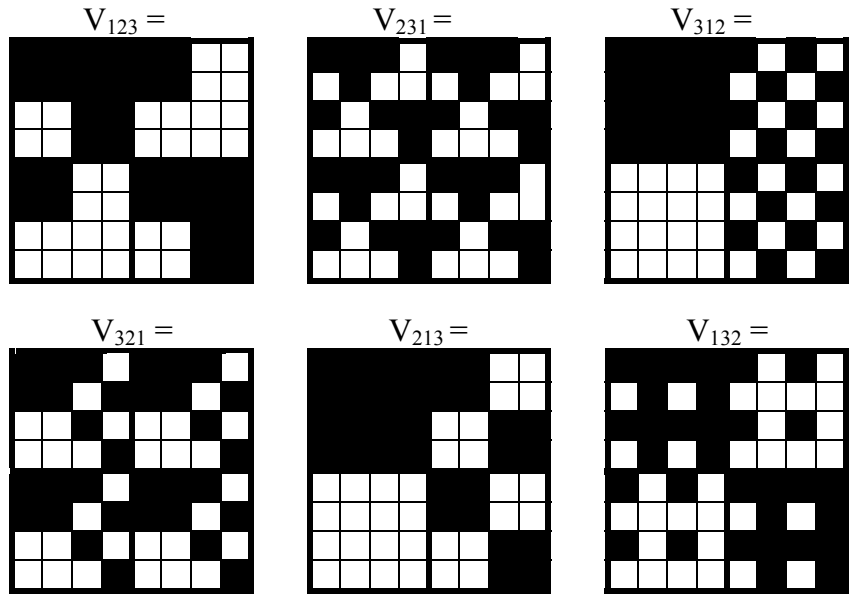

Figure 10. Numeric genomatrices $V_{123}$, $V_{231}$, $V_{312}$, $V_{321}$, $V_{213}$, $V_{132}$, in which each black cell means the element "+1" and each white cell means the element "-1".

By analogy with the expressions (1) we have again that matrices $W_{123} = 4^{-1}*V_{123}$, $W_{231} = 4^{-1}*V_{231}$, $W_{312} = 4^{-1}*V_{312}$, $W_{321} = 4^{-1}*V_{321}$, $W_{213,} = 4^{-1}*V_{213}$, $W_{132} = 4^{-1}*V_{132}$ are projection operators:

$$(W_{123})^2 = W_{123}; \quad (W_{231})^2 = W_{231}; \quad (W_{312})^2 = W_{312}$$

$$(W_{321})^2 = W_{321}; \quad (W_{213})^2 = W_{213}; \quad (W_{132})^2 = W_{132} \quad (2)$$

But what about alphabetic genomatrices with a non-diagonal disposition of complementary pairs C-G and A-U (Figure 11)? The third Kronecker power of each of such matrices forms a matrix of 64 triplets with an appropriate disposition of 32 black triplets and 32 white triplets. But this black-and-white mosaic in the 8*8-matrix does not correspond to any projection operator in contrast to the case of alphabetical genomatrices with a diagonal disposition of complementary pairs (Figure 8).

$$\begin{vmatrix} C & A \\ G & U \end{vmatrix}; \begin{vmatrix} C & U \\ G & A \end{vmatrix}; \begin{vmatrix} G & A \\ C & U \end{vmatrix}; \begin{vmatrix} G & U \\ C & A \end{vmatrix}; \begin{vmatrix} A & C \\ U & G \end{vmatrix}; \begin{vmatrix} A & G \\ U & C \end{vmatrix}; \begin{vmatrix} U & C \\ A & G \end{vmatrix}; \begin{vmatrix} U & G \\ A & C \end{vmatrix}$$

Figure 11. Examples of alphabetical genomatrices with a non-diagonal disposition of complementary pairs C-G and A-U.

### 4 About direct sums in biological structures

As we have mentioned in the section 2 of this article, projection operators are a powerful algebraic tool of studying important notion of a direct sum of vector spaces, etc. The Mendel's law on independent inheritance of traits allows thinking about every biological organism as a direct sum of such traits (in a frame of biological meanings of such inherited traits). The notion of direct sums can be used to study not only genetic sequences but also many natural and artificial systems which every person meets systematically: linguistic texts, musical compositions, architectural creations, etc. The described knowledge about genetic projection operators, which are used by nature in information bases of living matter, can lead to new

"genetic" approaches and methods in fields of analysis and algorithmic constructions of such natural and artificial systems. For example, on this way some famous ideas about structural connections between different oral languages and genetic languages which were put forward in works (Jacob, 1974, 1977; Jacobson, 1977, 1987, 1999; etc) can take new evidences; these ideas were discussed in a relation with matrix genetics in books (Petoukhov, 2008; Petoukhov, He, 2009).

Let us try to explain some thoughts, which arise here, by means of an example of musical compositions. People compose and listen to the music since ancient times. More than 30 thousand years ago, long before the appearance of arithmetic, our ancestors have played on stone flutes and on bone harps. Infants from the age of four months turned to the source of pleasant sounds (consonances) and turn away from unpleasant (dissonance). At the same time the human brain has not a specialized center for music, love for music can be assumed dispersed throughout the body (Weinberger, 2004). Many facts testify that musical feelings are based in some aspects on genetic mechanisms. Aristotle tried to understand how musical compositions (which contain only sounds) resemble the state of the soul. Leibniz considered the music as the unconscious exercise of soul in mathematics.

For a musical melody as a sequence of notes, a ratio of frequencies of its neighboring notes is important, but not the absolute values of the frequencies of individual notes. Related to this is that the melody is recognized, no matter in which the acoustic frequency range it is reproduced, for example, whether this melody is sung by male bass or by high-pitched voice of a woman or child. Concerning to durations of musical sounds, musical compositions use a certain set of relative durations which are named as "whole notes", "half notes", "quarter notes", "eighth notes", "sixteenth notes", "thirty-second note", "sixty-fourth notes".

Each monochromatic sound of a musical composition has its frequency (or its ratio to a basic frequency) and its relative duration. Such sound can be defined as a two-dimensional vector of a two-dimensional space which is a direct sum of two one-dimensional spaces (they play a role of coordinate axes of the plane): the one-dimensional space of relative frequencies and the one-dimensional space of relative durations of musical sounds. Let us take that "k" means a number of a musical sound in a sound sequence of the musical composition, a one-dimensional vector ($f_k$) means a relative frequency of this sound and a one-dimensional vector ($t_k$) means its relative duration. Then the first sound of the musical composition can be represented as a two-dimensional vector ($f_0$, $t_0$) on the base of the notion of a direct sum of vector spaces (in this example such order of components is used: $f_k$ stays before $t_k$).

Correspondingly the first four sounds of a musical composition can be represented as a 8-dimensional vector $X_0 = (f_0, t_0, f_1, t_1, f_2, t_2, f_3, t_3)$. The vector $X_0$ belongs to a 8-dimensional space which is a direct sum of the four two-dimensional spaces which contain vectors ($f_0$, $t_0$), ($f_1$, $t_1$), ($f_2$, $t_2$), ($f_3$, $t_3$). The next four musical sounds define a next 8-dimensional vector $X_1 = (f_4, t_4, f_5, t_5, f_6, t_6, f_7, t_7)$, etc. The whole musical composition can be represented as a sequence of such 8-dimensional vectors. Of course, the whole musical composition, which contains "n" sounds, can be represented also as a direct sum of "n" two-dimensional spaces of its separate sounds ($f_0$, $t_0$, $f_1$, $t_1$, ....., $f_n$, $t_n$) but we propose to pay a special attention to the case of 8-dimensional spaces which are connected with described (8*8)-genomatrices of 64 triplets.

In these "musical" vectors $X_0 = (f_0, t_0, f_1, t_1, f_2, t_2, f_3, t_3)$, $X_1 = (f_4, t_4, f_5, t_5, f_6, t_6, f_7, t_7)$, etc., relative frequencies $f_k$ of musical sounds occupy positions with even numbers k = 0, 2, 4, 6 and relative durations $t_k$ occupy positions with odd numbers k = 1, 3, 5, 7. Matrix genetics has revealed a connection of the genetic code with projection operators which transform the 8-dimensional vectors $X_0$, $X_1$, etc. into vectors where $f_k$ and $t_k$ are separated each from other. The speech is that some kinds of genoprojectors exist (see their examples **$f_0$** and **$m_1$** in Figure 12) which were revealed early in a course of studying a phenomenology of genetic Yin-Yang algebras (or bipolar algebras) and which transform an arbitrary 8-dimensional vector $X = [x_0; x_1; x_2; x_3; x_4; x_5; x_6; x_7]$ into new vectors $Z_k$ where only even or odd components of the vector X are presented separately. For example, **f₀**\*X = Z₀ = [x0; x0; x2; x2; x4; x4; x6; x6] and **m₁**\*X= Z1=[x1; x1; x3; x3; x5; x5; x7; x7].

$$\mathbf{f_0} = \begin{pmatrix} 1 & 0 & 0 & 0 & 0 & 0 & 0 & 0 \\ 1 & 0 & 0 & 0 & 0 & 0 & 0 & 0 \\ 0 & 0 & 1 & 0 & 0 & 0 & 0 & 0 \\ 0 & 0 & 1 & 0 & 0 & 0 & 0 & 0 \\ 0 & 0 & 0 & 0 & 1 & 0 & 0 & 0 \\ 0 & 0 & 0 & 0 & 1 & 0 & 0 & 0 \\ 0 & 0 & 0 & 0 & 0 & 0 & 1 & 0 \\ 0 & 0 & 0 & 0 & 0 & 0 & 1 & 0 \end{pmatrix} ; \quad \mathbf{m_1} = \begin{pmatrix} 0 & 1 & 0 & 0 & 0 & 0 & 0 & 0 \\ 0 & 1 & 0 & 0 & 0 & 0 & 0 & 0 \\ 0 & 0 & 0 & 1 & 0 & 0 & 0 & 0 \\ 0 & 0 & 0 & 1 & 0 & 0 & 0 & 0 \\ 0 & 0 & 0 & 0 & 0 & 1 & 0 & 0 \\ 0 & 0 & 0 & 0 & 0 & 1 & 0 & 0 \\ 0 & 0 & 0 & 0 & 0 & 0 & 0 & 1 \\ 0 & 0 & 0 & 0 & 0 & 0 & 0 & 1 \end{pmatrix}$$

Figure 12. Genetic projectors **f₀** and **m₁** from the genetic Yin-Yang algebra (or bipolar algebra) $YY_8$ which were described in our works (Petoukhov, arXiv:0803.3330 and arXiv:0805.4692; Petoukhov, He, 2009).

The knowledge about genetic projectors, about mathematical formalisms of vector spaces and about ideas of direct sums gives new approaches for analyzing and for algorithmic synthesis of musical compositions.

## 5 About other variants of genetic projectors

The described facts about close connections between projection operators and matrix presentations of the genetic code have put forward a task of a systematic study of all possible variants of genetic projectors which are based on matrix presentations of different characteristics of molecular ensembles of genetic systems. In this article we present only some of genetic projectors from their collection which is studied now.

Some examples of genetic projectors exist in a set of genetic bipolar algebras (Petoukhov, arXiv:0803.3330 and arXiv:0805.4692; Petoukhov, 2008; Petoukhov, He, 2009) where different matrix presentations of quasi-real units $f_0$ and $m_1$ satisfy the condition $f_0^2=f_0$ and $m_1^2=m_1$. In addition, quasi-real units of genetic bipolar algebras, which were constructed in a relation with a genetic matrix proposed by J.Kappraff and G.Adamson (Kappraff, Adamson, 2009; Kappraff, Petoukhov, 2009), are genoprojectors also.

Another kind of examples is connected with a presentation of oppositional complementary pairs C-G and A-U by means of oppositional signs: C=G=+1 and A=U=-1. Taking these numeric presentations of the genetic letters into account, one can present each triplet as a relevant product of these numbers "+1" and "-1". For instance in this numeric presentation the triplet CAG is equal to 1\*(-1)\*1=-1. Correspondingly the symbolic genomatrix of 64 triplets [C A; U G]$^{(3)}$ (Figure 2) is transformed in this case into the following numeric matrix (Figure 13):

$$Z = 8^{-1} * \begin{pmatrix} 1 & -1 & -1 & 1 & -1 & 1 & 1 & -1 \\ -1 & 1 & 1 & -1 & 1 & -1 & -1 & 1 \\ -1 & 1 & 1 & -1 & 1 & -1 & -1 & 1 \\ 1 & -1 & -1 & 1 & -1 & 1 & 1 & -1^* \\ -1 & 1 & 1 & -1 & 1 & -1 & -1 & 1 \\ 1 & -1 & -1 & 1 & -1 & 1 & 1 & -1 \\ 1 & -1 & -1 & 1 & -1 & 1 & 1 & -1 \\ -1 & 1 & 1 & -1 & 1 & -1 & -1 & 1 \end{pmatrix}$$

Figure 13. The genetic (8\*8)-projector on the base of the genetic matrix [C A; U G]$^{(3)}$ in the case of numeric presentations C=G=+1 and A=U=-1.

This genoprojector Z possesses interesting properties relative to 8-dimensional vectors, components of which form a sequence with uniform changes like the vector V=[x; x+y; x+2*y; x+3*y; x+4*y; x+5*y; x+6*y; x+7*y]. The action of genoprojector Z on the vector V leads to a vector with zero components: Z*V=[0;0;0;0;0;0;0;0]. The action of the complementary genoprojector (E-Z) on the same vector V does not change it: (E-Z)*V=V (here E is identity matrix). Components x, x+y, etc. of the vector V can be not only numbers but symbols also (letters, words, etc.)

## 8  Conclusion remarks

The Mendel's law of independent inheritance of traits demonstrates that molecular-genetic information at micro-levels defines traits on macro-levels of biological bodies. It testifies that biological organisms are some combinatorial algorithmic machines. Matrix genetics aims to reveal secrets of these algorithmic bio-machines.

Discrete character of the genetic code shows that genetic informatics is based on principles of combinatorics. Our discovery of close connections of projection operators with matrix presentation of structural phenomena of the genetic code clarifies that genetic combinatorics can be interpreted as a combinatorics of finite-dimensional discrete vector spaces. We propose using formalisms of genetic projectors and direct sums to study biological phenomena which are connected with genetic systems. The phenomenological facts of unexpected relations of genetic structures with special classes of projection operators, Hadamard matrices, Rademacher functions, etc. testify that the genetic code is an algebraic code in its nature. The results of matrix genetics allow developing new methods in bioinformatics and in applied fields of genetic algorithms. They are important also for developing algebraic biology, theoretical biology and for some other scientific and cultural fields which are related with genetic code systems.

**Acknowledgments**: Described researches were made by the author in the frame of a long-term cooperation between Russian and Hungarian Academies of Sciences and in the frame of programs of "International Society of Symmetry in Bioinformatics" (USA, http://polaris.nova.edu/MST/ISSB) and of "International Symmetry Association" (Hungary, http://symmetry.hu/). The author is grateful to Frolov K.V., Adamson G., Darvas G., Kappraff J., He M., Ne'eman Y., D.G.Pavlov, Vladimirov Y.S. for their support.